\documentclass[aps,prl,reprint, superscriptaddress,preprintnumbers, amsmath,amssymb,amsfonts, nofootinbib]{revtex4-1}
\usepackage{graphicx}
\usepackage[dvipsnames]{xcolor}
\usepackage{hyperref}
\usepackage{xspace}
\usepackage{ifdraft}
\usepackage{epstopdf}
\usepackage{slashed}
\usepackage{relsize}

\usepackage{tikz-feynman} 

\tikzfeynmanset{compat=1.1.0}
\tikzfeynmanset{/tikzfeynman/warn luatex = false}

\definecolor{jaxoblue}{HTML}{0086FF}

\definecolor{funRed}{RGB}{161,0,0}
\definecolor{funBlue}{RGB}{0,100,144}
\definecolor{cutred}{RGB}{219,56,49}
\definecolor{hgreen}{RGB}{25,176,146}
\definecolor{hgreen1}{RGB}{175,230,175}
\definecolor{hblue}{RGB}{52,152,219}
\definecolor{hblue1}{RGB}{255,255,166}
\definecolor{hred}{RGB}{216,83,117}
\definecolor{hred1}{RGB}{255,155,155}
\definecolor{cutred}{RGB}{219,56,49}
\definecolor{hgrey4}{RGB}{75,75,75}
\definecolor{hgrey5}{RGB}{50,50,50}
\definecolor{hgrey3}{RGB}{100,100,100}
\definecolor{hgrey}{RGB}{125,125,125}
\definecolor{hgrey2}{RGB}{125,125,125}
\definecolor{hgrey1}{RGB}{150,150,150}
\definecolor{hgrey0}{RGB}{175,175,175}
\definecolor{darkgreen}{RGB}{59,126,108}

\usepackage{xcolor}
\hypersetup{
    colorlinks,
    linkcolor={funRed},
    citecolor={funBlue},
    urlcolor={funBlue}
}

\newcommand{\fivegraphGluePropOp}[9]{ {
\begin{tikzpicture}[baseline=(current  bounding  box.center)]
\begin{feynman}
\vertex (a1) at (-2.4, -1) {\(\,#5\,\)};
\vertex (a2) at (-2.6, 0) {\(\,#6\,\)};
\vertex (a5) at (-2.4, 1) {\(\,#7\,\)};
\vertex [blob,#4](mid1) at (-1,0) {\(\,#1\,\)};
\vertex [blob,hgrey4](mid1x) at (-1,0) {$DDD\,\,\,$};
\vertex [blob,#4](mid1) at (-1,0) {$DDD\,\,$};
\vertex [white](mid3) at (-1,0) {\(\,#1\,\)};
\vertex [blob,hgrey4](mid2x) at (0.7,0) {$DD\,$};
\vertex [blob,#3](mid2) at (0.7,0) {\(\,#2\,\)};
\vertex [white](mid4) at (0.7,0) {\(\,#2\,\)};
\vertex (a3) at (1.8, 1) {\(\,#8\,\)};
\vertex (a4) at (1.8, -1) {\(\,#9\,\)};
\vertex (a7) at (0, -0.9);
\vertex (a6) at (0, 0.9);
\diagram{
(mid2x) --[gluon, thick](a3),
(mid2x) --[gluon, thick](a4),
(mid2x) -- [gluon, thick](mid1x),
(a1)--[gluon, thick](mid1x),
(a2) -- [gluon, thick](mid1x),
(mid1x) --[gluon, thick](a5),
(a6)--[scalar, ultra thick,cutred](a7)
};
\end{feynman}
\end{tikzpicture}
}
}

 \def\draftnote#1{{\color{red}\it #1}} 

\def\fig#1{Fig.~{\ref{#1}}}

\def\spa#1.#2{\left\langle#1\,#2\right\rangle}
\def\spb#1.#2{\left[#1\,#2\right]}
\def\spash#1.#2{\spa{\smash{#1}}.{\smash{#2}}}
\def\spbsh#1.#2{\spb{\smash{#1}}.{\smash{#2}}}
\def\sand#1.#2.#3{%
\left\langle\smash{#1}{\vphantom1}^{-}\right|{#2}%
\left|\smash{#3}{\vphantom1}^{-}\right\rangle}
\def\sandpp#1.#2.#3{%
\left\langle\smash{#1}{\vphantom1}^{+}\right|{#2}%
\left|\smash{#3}{\vphantom1}^{+}\right\rangle}
\def\sandpm#1.#2.#3{%
\left\langle\smash{#1}{\vphantom1}^{+}\right|{#2}%
\left|\smash{#3}{\vphantom1}^{-}\right\rangle}
\def\sandmp#1.#2.#3{%
\left\langle\smash{#1}{\vphantom1}^{-}\right|{#2}%
\left|\smash{#3}{\vphantom1}^{+}\right\rangle}

\def\Tr{\, {\rm Tr}}

\def\eqn#1{eq.~(\ref{#1})}

\def\eqns#1#2{eqs.~(\ref{#1}) and~(\ref{#2})}

\def\be{\begin{equation}}
\def\ee{\end{equation}}
\def\bea{\begin{eqnarray}}
\def\eea{\end{eqnarray}}
\def\ba{\begin{eqnarray}}
\def\ea{\end{eqnarray}}

 \definecolor{MattOrange}{rgb}{1.0,0.4,0.2}

\newcommand{\Lfth}{\ensuremath{\mathcal{L}_{\text{YM}+F^3}}}

\newcommand{\tfth}{ \ensuremath{\Tr(F^3)}}

\begin{document}

\preprint{
}

\author{John Joseph M. Carrasco}
\affiliation{Department of Physics and Astronomy, Northwestern
  University, Evanston, Illinois 60208, USA}
\affiliation{Institut de Physique Th\'{e}orique, Universite Paris Saclay, CEA, CNRS, F-91191 Gif-sur-Yvette, France}
\author{Matthew Lewandowski}
\affiliation{Department of Physics and Astronomy, Northwestern
  University, Evanston, Illinois 60208, USA}
\author{Nicolas H. Pavao}
\affiliation{Department of Physics and Astronomy, Northwestern
  University, Evanston, Illinois 60208, USA}

\title{The Color-Dual Fates of $F^3$, $R^3$, and $\mathcal{N}=4$ Supergravity}

\begin{abstract}
We find that the duality between color and kinematics can be used to inform the high energy behavior of effective field theories. Namely, we demonstrate that the massless gauge theory of Yang-Mills deformed by a higher-derivative $F^3$ operator cannot be tree-level color-dual while consistently factorizing without a tower of additional four-point counterterms with rigidly fixed Wilson coefficients that reaches to the ultraviolet (UV).  We find through explicit calculation a suggestive resummation, namely that their amplitudes are consistent with the $\alpha'$ expansion of those generated by the $(DF)^2+\text{YM}$ theory, a known color-dual theory where the $F^2$ term has been given a mass squared proportional to $1 / \alpha'$.  As a result, considering consistent double-copy construction as a physical principle implies that an  $F^3$-based color-dual resolution of the UV divergence in $\mathcal{N}=4$ supergravity comes at the cost of field-theoretic locality.  Similarly, when double-copying $F^3$ with itself,  double-copy consistency lifts $R^3$ gravity to a family of gravity theories with an all-order tower of higher-derivative corrections, which includes the closed bosonic string as a standard adjoint-type double-copy.
\end{abstract}

\maketitle

\section{Introduction}

Perturbative calculation in quantum gravity theories is not as prohibitive as Feynman diagram approaches, in generic gauges, might suggest. The duality between color and kinematics~\cite{BCJ}, and associated double-copy construction~\cite{KLT,BCJ,BCJLoop}, reduce the complexity of calculations in many gravity theories to understanding predictions in much simpler gauge theories. 

On the other hand, identifying consistent ultraviolet (UV) completions of quantum gravity theories still remains challenging from the perspective of point-like quantum field theories. The only proven UV completion to quantum gravity, the closed string, requires an infinite number of higher-derivative corrections from the QFT perspective--arguably rendering the theory non-local. The best candidate for a perturbatively finite local quantum field theory of gravity is the maximally supersymmetric theory~\cite{Cremmer:1978km}, $\mathcal{N}=8$ supergravity (SG), which remains finite in four dimensions at least through the five-loop correction~\cite{GravityThree,GravityFour,UVFiveLoops}. Counterterms compatible with known linearly-realized symmetries have been identified which would be relevant starting at seven loops~\cite{Beisert2010jx,Bossard:2011tq}, although their coefficients have not been determined and could vanish in four dimensions. The ultimate fate of $\mathcal{N}=8$ SG awaits explicit calculation.   

  Absent direct data from the UV theory, positivity bounds have long\footnote{See, ref.~\cite{Adams:2006sv} and refs.~\cite{deRham:2017avq,Bellazzini:2019xts,Bellazzini:2020cot,Bern:2021ppb,Creminelli:2022onn} for recent examples.} been a tool for probing potentially valid UV behavior of effective field theories (EFTs) by bounding  \textit{a priori} unconstrained Wilson coefficients. In this Letter we investigate whether the duality between color and kinematics in combination with factorization constraints can serve a similar role and go beyond simply aiding in calculation.  After all, perturbative string theory, the only known UV completion to Einstein-Hilbert gravity, can now be understood at tree-level as a field-theoretic double-copy involving all-order in $\alpha'$  color-dual EFTs~\cite{MafraBCJAmplString,Broedel:2013tta,Carrasco:2016ldy,Mafra:2016mcc,Carrasco:2016ygv}. We demonstrate that this duality can indeed inform UV completion, finding surprisingly that it has the potential to enforce all-order relations between Wilson coefficients starting only from the IR. 
We motivate by engaging with a sharp problem -- resolving the UV behavior of half-maximal supergravity in four-dimensions.

In four-dimensions, the  perturbative finiteness of pure half-maximal SG survives three-loops~\cite{Bern:2012cd}, a challenge analogous to maximal supergravity's conjectured seven-loop divergence~\cite{Bossard:2011tq}, only to diverge at four loops~\cite{N4GravFourLoop}. The observed divergence at four loops has been linked to the $U(1)$ anomalous behavior~\cite{MarcusAnomaly,CarrascoN4Anomaly,Bern:2017tuc} of the theory.  
Such anomalous behavior at one loop can be removed with a simple local counterterm whose double-copy description involves adding the $\tfth$ operator to a pure Yang-Mills theory.  Does including this counterterm render supergravity finite?  This too awaits explicit calculation -- but investigation at one and two loops~\cite{BPRAnomalyCancel,Bern:2019isl} has verified that the addition of this appropriately tuned counterterm does indeed remove the anomalous behavior.

We will show in this Letter that the consequence of requiring that amplitudes both consistently factorize and participate in double-copy construction via color-kinematics duality, a property we call \textit{double-copy consistent}, demands rigid relations between coefficients of EFT operators that ascend into the UV. Motivated by the 4D anomaly of half-maximal supergravity, we will investigate amplitudes in a Yang-Mills theory deformed by the $\tfth$ operator. We will do so in $D$-dimensions using formal polarization vectors. For a detailed review of double-copy structure and supersymmetry, we refer the interested reader to ref.~\cite{Bern:2019prr}. Here we need only recall that the double-copy structure of pure half-maximal supergravity (half-max.~SG) is given,
\begin{equation}
  \left( \text{half~max. SG} \right) \, =\, \left(\text{maximal sYM} \right) \otimes \text{YM}\,.
\end{equation}
Maximally supersymmetric Yang-Mills (maximal sYM) follows via dimensional reduction of one supersymmetry in ten dimensions. The double-copy of maximal sYM in any dimension with non-supersymmetric gauge theory results in a supergravity with half the maximal supersymmetry it could have in that dimension, so e.g. $\mathcal{N}=4$ SG in 4D.

First, we provide evidence that double-copy consistency for $\text{YM}+F^3$ requires the inclusion of an infinite tower of rigidly constrained counterterms at four-points through $\mathcal{O}(\alpha'^4)$ by explicit calculation via color-dual bootstrap between four and five-points--finding for the first time the most generic color-dual five-vector amplitude through this mass dimension.  We parameterize all residual freedom in Wilson coefficients  unconstrained by five-point factorization. We then present a potential resummation of this tower of operators to 
$(DF)^2+\text{YM}$ theory, a known dimension-six color-dual theory, whose $\alpha'$ expansion explicitly matches the Wilson 
coefficients that result from our bootstrap.  

What has happened to our half-maximal supergravity? Remarkably, we have bootstrapped to a string theory where some (but not all) of the non-locality has been removed -- the double-copy of $(DF)^2+\text{YM}$ with maximal sYM results in amplitudes of a twisted\footnote{In twisted string theories~\cite{Hohm:2013jaa,Huang:2016bdd} a relative sign flip of the inverse string tension between the holomorphic and anti-holomorphic sectors results in a finite physical spectrum.} heterotic string~\cite{Guillen:2021mwp}. This is an entirely novel consequence of demanding color-dual consistency from an IR vector theory starting only with Yang-Mills deformed by $F^3$.   We note that there remains the color-dual freedom to complete to known UV completions like the standard heterotic string. We clarify this structure and freedom, while pointing out consequences for double-copy consistent gravity theories involving $R^3$.  We close this Letter by summarizing our results and discussing important next-steps. 

%
\section{Double-copy consistency of $\tfth$}

The idea of double-copy consistency gets to the heart of an open question regarding double-copy construction.  Should we regard the double  copy as a technical trick to be employed piecemeal, amplitude by amplitude as necessary, or rather as a physical principle pointing to the presence of an as-yet-unrecognized physical mechanism braiding together factors of two otherwise consistent theories?  Here we explore the gravitational UV consequences of using double-copy consistency to constrain an ansatz-driven color-dual vector bootstrap. 

For comparison, we first consider a scalar EFT that also requires an infinite number of counterterms to be double-copy consistent.  We start with a theory of massless scalars with only the interaction term,
\begin{equation}
   \mathcal{L}_{4\text{-int}}=  \Lambda  f^{abe} f^{ecd} (\partial_\mu \phi_a) \phi_b \phi_c (\partial^\mu \phi_d)\,.
   \label{eqn:scalarExample}
\end{equation}
 While even-multiplicity amplitudes are non-vanishing, the four-point amplitude is color-dual.  The color-dual theory requires an additional six-field operator whose coefficient is uniquely determined by the duality and consistent factorization~\cite{Carrasco:2019qwr}. Indeed requiring double-copy consistency to arbitrary multiplicity involves adding an infinite chain of operators with fixed Wilson coefficients. This resums to the pion Lagrangian of the venerable nonlinear sigma model known to be color-dual~\cite{Cheung2016prv}.  Demanding double-copy consistency encodes the same physical Nambu-Goldstone symmetry as imposing the constraint of the famous Adler's zero.

Let us now address the theory at the heart of this Letter, Yang-Mills theory deformed by the higher derivative $\tfth$ operator,
\begin{equation}
\label{YMF3Lagrangian}
    \Lfth= -\frac{1}{4}\Tr(F^2)+\frac{\alpha'}{3}  \tfth\,.
\end{equation}
For this theory both $\mathcal{O}(\alpha'^0)$ and $\mathcal{O}(\alpha'^1)$  orders in the three-gluon amplitude satisfy the duality between color and kinematics,
as does the four-point tree-level amplitude through $\mathcal{O}(\alpha')$~\cite{Broedel2012rc}.   In contrast, the $\mathcal{O}(\alpha'^{\,2})$ contribution to the four-point amplitude, while naturally gauge invariant, is not color-dual.  One must modify $\Lfth$ with an additional $\Tr(F^4)$ operator~\cite{Broedel2012rc} for color-dual four-point amplitudes through $\mathcal{O}(\alpha'^{\,2})$. 

In fact, we will present evidence that \textit{no finite number} of local operators is sufficient to render $\Lfth$ double-copy consistent. The situation is markedly different than the earlier scalar theory, which at least requires only a finite number of operators to render any particular multiplicity color-dual. 

\begin{figure}[t]
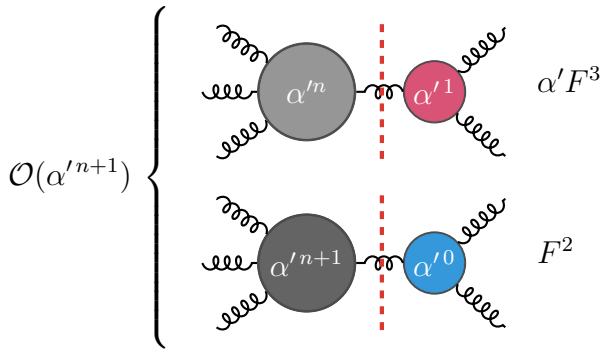

\larger[2.4]
\[
\mathcal{O}(\alpha'^{\,n+1})\,\begin{cases}
\fivegraphGluePropOp{\alpha'^{n}}{\alpha'^{\,1}}{hred}{hgrey1}{}{}{}{}{}& \!\!\!\!\! {\alpha' F^3}
\\
\fivegraphGluePropOp{\,\alpha'^{\,n+1}\,}{\alpha'^{\,0}}{hblue}{hgrey3}{}{}{}{}{}& \!\!\!\!\! {F^2}
 \end{cases}
 \]
 \smaller[2.4]
 \caption{Contributions to the factorization of five-point tree-level amplitude, \eqn{cutFive}, at $\mathcal{O}(\alpha'^{n})$. Color-dual constraints on the five-point amplitude relate Wilson coefficients of $\alpha'^n$ and $\alpha'^{n+1}$ four-field operators.}
\label{fig:fiveFactor}
\end{figure}

To understand the origin of the requisite UV ladder, consider the minimal cut factorization of five-points,
\begin{equation}
    \mathcal{A}_5 {(12345 )} |_{(k_4+k_5)^2\text{-cut}}=\sum_{\text{states}} \mathcal{A}_4(123l^s) \mathcal{A}_3(-l^{\overline{s}}45)\,.
\label{cutFive}
\end{equation}
Since the three-point amplitude is comprised of both Yang-Mills and $\tfth$ contributions, {$A_3   \equiv  A_3^{\text{YM}}+\alpha' A_3^{F^3}$}, the physical state-sum with a purely local four-point contribution at order $\mathcal{O}(\alpha'^{\,n})$ means a non-vanishing five-point factorization channel at $\mathcal{O}(\alpha'^{\,n+1})$. We find that the $\mathcal{O}(\alpha'^{\,n+1})$ contribution at five-points cannot be color-dual without the additional contribution from a specific four-field operator of order $\mathcal{O}(\alpha'^{\,n+1})$ sewn with the $\text{Tr}(F^2)$ term, as depicted in \fig{fig:fiveFactor}. This higher-weight four-point contact must come in with a fixed Wilson coefficient. Contraction of this additional $\mathcal{O}(\alpha'^{\,n+1})$ four-point contact with the $\alpha' \tfth$ term now forces consideration of a non-vanishing $\mathcal{O}(\alpha'^{\,n+2})$ contribution at five-points, and so forth, thereby constructing a compulsory ladder of operators into the UV.


We will now demonstrate the inevitability of this ladder through $\mathcal{O}(\alpha'^{\,4})$ -- a nontrivial task given the formidable size of the necessary five-vector ansatze. To tease out this structure, we start by identifying the color-dual four-point kinematics that are consistent  three-points, and then we calcualte the constraints of five-point factorization. 

At four-points adjoint color-dual amplitudes can always be written in terms of cubic, or trivalent, graphs:
\begin{equation}
	\mathcal{A}_4 = \frac{n_s c_s}{s} + \frac{n_t c_t}{t} + \frac{n_u c_u}{u}.
\end{equation}
We label the graphs by their propagators, using $s$, $t$ and $u$ to refer to the standard Mandelstam invariants, using an all outgoing convention, $s=(k_1+k_2)^2$, $t=(k_2+k_3)^2$, and $u=(k_1+k_3)^2$. Each graph has a color weight arising from dressing every vertex with a color-structure constant, e.g. $c_s=c(1,2,3,4)=f^{a_1a_2b}f^{ba_3a_4}$. Similarly, keeping polarization vectors formal, we require a functional map from the labeled graph to kinematic weight such that, e.g., $n_s=n(1,2,3,4)$, $n_t=n(1,4,3,2)$, and $n_u=n(1,3,2,4)$.
The dressing is color-dual if the kinematics satisfy a Jacobi-like relation $n_s=n_t+n_u$ and are antisymmetric about vertex flips in concordance with the adjoint color-factors, e.g.~$n(1,2,3,4)=-n(2,1,3,4)$. Only if such a representation exists can the theory participate in double-copy construction.

We can bootstrap to arbitrary orders in $\alpha'$ by giving our color-dual numerator a generic ansatz in terms of $D$-dimensional formal polarization vectors at the particular mass-dimension of interest, and then constraining on any factorization channels.  Here we have complete control over color-dual structure in terms of only eight vector building blocks that span all higher-derivative corrections under composition with scalar permutation invariants~\cite{Carrasco:2019yyn,Carrasco:2021ptp}. We simply impose factorization to three-point vertices of the form $A^{\text{YM}}_3 + \alpha' A^{F^3}_3$. As such, every contribution to $\mathcal{A}_4$ above $\mathcal{O}(\alpha'^{\,2})$ must be purely local. All lower-order terms are fixed to be related to coupling constants appearing in the three-point amplitude.

We find, through $\mathcal{O}(\alpha'^4)$, the following functional numerator of the $s$-channel graph, 
\be
\begin{split}
\label{eq:4pointnum}
&{n}^{\text{dcc}}_s = n_s^{\text{YM}}  + \alpha' n_s^{\text{YM}+F^3} +  \alpha'^{\,2}\, n_s^{(F^3)^2+F^4}+ \\
          & \alpha'^{\,3} \left [ a_{3} \, \left(  n_s^{D^2F^4}  +  \sigma_2 \, n_s^{\text{YM}+F^3}  \right)+ a_{3,\text{YM}} \,\sigma_3\,  n_s^{\text{YM}} \right] + \\
           &\alpha'^{\,4} \left[ a_{4,1} \, \left(  n_s^{(DF)^4_1}+  \sigma_2 \, n_s^{(F^3)^2+F^4}  \right ) \right.+ \\
  & \qquad \qquad\!\! \left. a_{4,2} \, n_s^{(DF)^4_2} + a_{4,F^3}\,\sigma_3 \,n_s^{\text{YM}+F^3} \right] + 
   \mathcal{O}(\alpha'^5)\,.  
\end{split}
\ee
 We have scaled out mass-dimension using the $\tfth$ dimensionful coupling $\alpha'$, leaving all unconstrained ansatz parameters, $a_i$, dimensionless.  We introduce scalar permutation invariants $\sigma_2 =  (s^2+t^2+u^2)/8$ and $\sigma_3 = (s t u)/8$. Here the numerator is given in terms of six of the eight spanning color-dual vector building blocks $n_s^{\mathcal{O}_i}$ of ~\cite{Carrasco:2019yyn,Carrasco:2021ptp}, with explicit definitions given in an ancillary machine-readable file.   The factors of $a_i$ are free numeric parameters unconstrained by the factorization of four-points.  We will now see that imposing factorization constraints on the most general color-dual five-point amplitude entirely fixes $a_{3}$, $a_{4,1}$, and $a_{4,2}$,  and relates $a_{4,F^3}$ to $a_{3,\text{YM}}$.

As no spanning basis of color-dual vector building
blocks is known yet at five-points, we consider a general ansatz 
expressed in all combinations of Lorentz invariants. Performing the calculation to the desired order in mass-dimension involves reduction of a 58,923 parameter ansatz spanning from order four in dot products, relevant to $\mathcal{O}(\alpha'^0)$, to order eight at mass-dimension $\mathcal{O}(\alpha'^{\,4})$ above Yang-Mills.  Unsurprisingly reaching this order via vector ansatze is a computational task comparable to nontrivial multi-loop calculations only recently within reach~\cite{Dixon:2022rse}.

We find that requiring color-dual factorizing amplitudes at five-points through $\mathcal{O}(\alpha'^4)$ imposes the following constraints on the free parameters in \eqn{eq:4pointnum},
\begin{equation}
	a_{3}=a_{4,1}=a_{4,2}=1 \,,\qquad 	
	a_{4,F^3}= 1+a_{3,\text{YM}} \,.
\end{equation}
In summary, double-copy consistency establishes a tower of ever higher-derivative operators at four-points with rigidly locked Wilson coefficients. The tower remains to be proven to all orders in $\alpha'$, but we find the explicit results so far to be sufficiently provocative to consider resummation.

\section{Resummation to $DF^2+\text{YM}$}
A natural question is if the results above resum to a known theory.  We apparently have the freedom to set $a_{3,\text{YM}}=0$. If we do so,  our four-point and five-point amplitudes precisely match the $\mathcal{O}(\alpha'^4)$ expansion of the $B(1,\ldots,n)$ amplitudes of \cite{Huang:2016tag}.   These $B$ amplitudes belong~\cite{Azevedo:2018dgo} to the $(DF)^2+\text{YM}$ theory of ref.~\cite{Johansson:2017srf}, where the $(DF)^2$ has been deformed by a massive gauge-theory, with mass scale set by $1/\alpha'$. Indeed, in ref.~\cite{Carrasco:2019yyn}, the four-point amplitude of $(DF)^2+\text{YM}$ theory was expressed in terms of the above color-dual building blocks: 
\begin{multline}
	n_s^{(DF)^2+\text{YM}} = n_s^{\text{YM}}+\\
	 \frac{\alpha' n_s^{F^3}+\alpha'^{\,2} n_s^{(F^3)^2+F^4} + \alpha'^{\,3} n_s^{D^2F^4} + \alpha'^{\,4} n_s^{(DF)^4}}{1-\alpha'{}^2\, \sigma_2 - \alpha'{}^3 \, \sigma_3} \ , \label{resummedDF} 
 \end{multline} with $n_s^{(DF)^4}\equiv n_s^{(DF)^4_1}+n_s^{(DF)^4_2}$.

 The $(DF)^2+\text{YM}$ theory that generates the $B$ amplitudes is a fascinating color-dual dimension-six theory involving the $\tfth$ operator with higher-order propagators. It was first written down by Johansson and Nohle~\cite{Johansson:2017srf} with the explicit aim of finding a double-copy description of conformal supergravity. While we double-copy over standard propagators, the hallmark conformal propagators emerge from the fact that the resummed graph ``numerators'' of \eqn{resummedDF} are themselves non-local.

As previously noted, double-copying $(DF)^2+\text{YM}$ with maximal sYM recovers the four graviton amplitude of the twisted heterotic string~\cite{Guillen:2021mwp}. More generally here we find ourselves lifting the Poincar\'e theory to a family of Einstein-Weyl theories of which Berkovits-Witten conformal supergravity is a famous limiting example~\cite{Berkovits:2004jj,JohanssonConformal},
\begin{multline}
	(\text{half-max. Einstein-Weyl} + \ldots )  = \\
	\left(\text{maximal sYM}\right)  \otimes \left( (DF)^2+\text{YM} + \ldots \right)\,. 
\end{multline}
We include ellipses to emphasize the potential inclusion of  operators unfixed by solely requiring the double-copy consistency of $\text{YM}+F^3$.   As we discuss in the next section, this freedom can be fixed with particular Wilson coefficients to promote half-maximal SG amplitudes to the gravitational amplitudes of the heterotic string at tree-level. 

Using \eqn{resummedDF} to rewrite our constrained ansatz in \eqn{eq:4pointnum},  offers a revised form of the four-point numerator for our double-copy consistent theory through $\alpha'^4$: 
\begin{equation}
n^{\text{dcc}} = n^{(DF)^2+\text{YM}} +a_{3,\text{YM}}\, \alpha'{}^3\,\sigma_3 \left(n^{\text{YM}}+ \alpha' n^{F^3}  \right)
\label{eqnFreedom}
\end{equation}
Note that the terms in the second expression, $n^{\text{YM}}+\alpha' n^{F^3}$, mirror the first terms of the $\alpha'$ expansion of $n^{(DF)^2+\text{YM}}$ given in \eqn{resummedDF}. This suggests the possibility that double-copy consistent amplitudes can be promoted to higher-order contact terms via a product of their color-dual numerators with scalar permutation invariants, and that this information can be consistently propagated to higher multiplicity color-dual amplitudes. 

Indeed, we will shortly introduce a map from string-theory which offers not only a proof of concept, but a prescriptive understanding of how higher multiplicity color-dual amplitudes may be constructively reconciled with the addition of local counterterms. It is therefore likely that the we span all order contributions to a double-copy consistent$\tfth$-theory four-vector amplitude with,
\begin{equation}
\label{eq:scc4point}
A^{\text{dcc}}_4 = B(1,2,3,4) \left[1+\!\!\sum_{x\geq 1,y}c_{(x,y)}\sigma_3^x\sigma_2^y\alpha'^{\,3x+2y}\right]\, ,
\end{equation}
where  $\sigma_3$ and $\sigma_2$ are the four-point scalar permutation invariants, and all remaining freedom is parameterized by $c_{(x,y)}$, which encode the Wilson coefficients of higher-derivative corrections. Higher-multiplicity factorization may yet require additional relations between $c_{(x,y)}$, but none that could exclude the single-valued promotion of \eqn{defSV} that we will now describe.

%
%
\section{Heterotic string and the SV promotion}

We now demonstrate that the additional UV freedom to add operators to the $(DF)^2+\text{YM}$ theory allows us to promote the half-maximal Einstein-Weyl supergravity amplitudes to the tree-level graviton amplitudes of the heterotic string. Recall that ordered open superstring amplitudes emerge from the field theory double-copy of Yang-Mills with doubly-ordered $Z$-theory amplitudes~\cite{MafraBCJAmplString,Broedel:2013tta,Carrasco:2016ldy,Mafra:2016mcc,Carrasco:2016ygv}:
\begin{equation}
\label{eq:kltOSS}
    A^{\text{OSS}}_A =  A^{\text{sYM}}_a \otimes^{ab}  Z_{Ab}\, ,
\end{equation}
where the indices, $a, b$ and $A$, refer to various orderings of kinematic labels, and the outer product is taken to mean the field-theoretic double-copy.

The doubly-ordered scalar $Z$-theory disc amplitudes encode string-theoretic higher-derivative corrections at each order in $\alpha'$. Here we use capital indices to refer to orderings of external legs that satisfy string-theoretic monodromy relations~\cite{Monodromy,Stieberger:2009hq}, and the lowercase indices to refer to orderings that satisfy field-theoretic amplitude relations~\cite{KleissKuijf,BCJ}.   This notation emphasizes the fact that the bi-color dressed  Z-theory amplitudes have the property that their $a$-ordered amplitudes are color-dual order by order in $\alpha'$~\cite{Carrasco:2019yyn, Carrasco:2021ptp}.

A similar double-copy structure exists for the open bosonic string~\cite{Huang:2016tag,Azevedo:2018dgo}, 
\begin{equation}
    A^{\text{OBS}}_A = B_a\otimes^{ab}  Z_{Ab}\, ,
    \label{eq:kltOBS}
\end{equation}
where we again use $B$ to refer to ordered amplitudes generated by the $(DF)^2+\text{YM}$ theory.

Closed superstrings are also field-theoretic double-copies to all multiplicity.  This can be seen by first noting the construction of
closed string amplitudes via the string KLT kernel~\cite{KLT}, represented here by $\otimes_{\alpha'}$,
\begin{equation}
    A^{\text{CSS}}= A^{\text{OSS}}_{A} \otimes^{AB}_{\alpha'} A^{\text{OSS}}_B \ . 
\end{equation}
Applying now \eqn{eq:kltOSS} reveals a field-theory double-copy,
\begin{align}
    A^{\text{CSS}}&= (A^{\text{sYM}}_a \otimes^{ab}Z_{Ab}) \otimes^{AB}_{\alpha'} (Z_{Bc}\otimes^{cd} A^{\text{sYM}}_d)\,\\
	   &= A^{\text{sYM}}_a \otimes^{ab} (A^{\text{sYM}})^{\text{sv}}_b \ , 
	\label{CSS}
\end{align}
where we introduce the \textit{single-valued promotion} of field-theory amplitudes,
\begin{equation}
	(Y)^{\text{sv}}_a \equiv (Z_{Aa} \otimes^{AB}_{\alpha'} Z_{Bb}\otimes^{bc} Y_c) .
	\label{defSV}
\end{equation}
This operation is called `single-valued' because all the coefficients of $\alpha'$ introduced by the promotion come with only single-valued multiple zeta values.  At four-points this can be understood as multiplying the $Y$ theory color-dual numerators by scalar permutation invariants at each order in $\alpha'$.  The existence of such a double-copy consistent map means that we are free to conjecture the most general double-copy consistent UV completion of $B(1,2,3,4)$ to be contained in \eqn{eq:scc4point}. 

It was pointed out in ref.~\cite{Azevedo:2018dgo} that amplitudes of $(DF)^2+\text{YM}$ theory also play a  critical role in the field-theoretic construction of gravitational heterotic string amplitudes,
\begin{equation}
    A^{\text{HS}}=B_a \otimes^{ab} \left(A^{\text{sYM}}\right)^{\text{sv}}_b\,.
\end{equation}
It is clear from the above construction that one could equally well describe the hetoretic string amplitude as
\begin{equation}
    A^{\text{HS}}=\left( B\right)^{\text{sv}}_a \otimes^{ab} A^{\text{sYM}}_b\,.
\end{equation}
The set of consistent double-copy completions to $\text{YM}+F^3$ must therefore allow for the single-valued promotion of $(DF)^2+\text{YM}$.  Indeed this is realized through $\mathcal{O}(\alpha'^4)$ by setting $a_{3,\text{YM}}=c_{(1,0)}=\zeta_3$ in \eqns{eqnFreedom}{eq:scc4point}.

It is natural at this stage to remark on the double-copy of $\text{YM}+F^3$ with itself. Do we generate $R^3$ from double-copy in the sense of the $\alpha'^2$ corrections to the closed string? The answer, as initially noted in ref.~\cite{Broedel2012rc}, is yes  gravitational amplitudes involving single insertions of $R^3$ do arise from amplitudes involving single insertions of $F^3$ double-copied with themselves. 
Following the analysis of this Letter, double-copy consistency lifts the result to a family of gravitational theories that includes~\cite{Azevedo:2018dgo} the tree-level amplitudes of the closed bosonic string, 
\begin{equation}
    A^{\text{CBS}} =  \left( B \right)^{\text{sv}}_a \otimes^{a b} B_b \,.
\end{equation}

\section{Conclusion}
We have presented evidence that demanding double-copy consistency of a gauge theory with the $\tfth$ operator induces an all-order tower of $\alpha'$ corrections, which seems to require at a minimum all higher-derivative corrections associated with $(DF)^2+\text{YM}$.   There exists a small basis of color-dual vector building blocks, up to trivial scalar permutation invariants, at four-points~\cite{Carrasco:2019yyn}.  Using this basis reduces the complexity of four-point color-dual vector amplitudes to simple considerations of what permutation invariant scalars are required for a given mass-dimension. Developing a similar basis for vector building-blocks at five-points, as has already been done for higher-derivative color-weights~\cite{Carrasco:2021ptp}, would allow a simple proof that $\text{YM}+F^3$ must close to $(DF)^2+\text{YM}$ under double-copy consistency.

Here we summarize the most important consequence of our analysis. If we require double-copy consistency as a matter of principle, and we wish to grapple with the UV behavior of half-maximal supergravity by adding the $\tfth$ operator to the Yang-Mills copy, it appears that the fate of the theory lies in a family of Einstein-Weyl theories with freedom to add an additional tower of higher-derivative corrections compatible with the single-valued promotion. Our findings invite a new paradigm that elevates color-kinematics duality from a mathematical correspondence with the capacity to encode IR symmetries like Adler's zero, to a principle capable of probing UV physics captured by higher-derivative corrections consistent with the heterotic string. 

Furthermore we identified the number-theoretic single-valued promotion of \eqn{defSV} as a tool for lifting double-copy consistent field theories to all orders in higher dimensional operators.  It will be intriguing to learn in what ways the fixed single-valued MZV Wilson coefficients of this mapping can be generalized to identify distinct classes of double-copy consistent theories and what principles from a field theory perspective uniquely select the single-valued promotion.

Finally, it does not escape us that our analysis has implications beyond half-maximal supergravity. Adding supersymmetric matter to the Yang-Mills single-copy would lift $\mathcal{N}=4$ SG to higher supersymmetry, evading known anomalies. We expect the explicit calculation of the UV behavior of $\mathcal{N}=5$ SG at five-loops and $\mathcal{N}=8$ SG at seven and eight-loops to prove critical to understanding the potential perturbative finiteness of four-dimensional local theories of gravity.

\begin{acknowledgements}
We gratefully acknowledge related collaboration and insightful conversations with Zvi Bern, James Mangan, Laurentiu Rodina, Radu Roiban, Aslan Seifi, and Suna Zekio\u{g}lu. We  would like to thank Henrik Johansson, James Mangan, Oliver Schlotterer, and Suna Zekio\u{g}lu for very helpful feedback regarding earlier drafts.  This work was supported by the DOE under contract DE-SC0021485, DE-SC0015910, and by the Alfred P. Sloan Foundation. M.L.~and~N.H.P. would additionally like to acknowledge the Northwestern University Amplitudes and Insight group,  Department of Physics and Astronomy, and Weinberg College for support. 
\end{acknowledgements}

\bibliography{Refs_f3ladder.bib}

\end{document}